\documentclass{elsarticle}

\usepackage{graphicx}
\usepackage{amssymb}
\journal{Commun. in Nonlin. Sci. and Num. Sim.}

\def\unsurn{\frac{1}{N}}

\def\undemi{\frac{1}{2}}

\begin{document}

\begin{frontmatter}
 \title{Numerical resolution of the Vlasov equation for the Hamiltonian Mean-Field model}
 \author{Pierre de Buyl}
 \address{Center for Nonlinear Phenomena and Complex Systems \\
 Universit{\'e} Libre de Bruxelles (U.L.B.), Code Postal 231, Campus Plaine, B-1050 Brussels, Belgium}  
 \begin{abstract}
   We present in this paper detailed numerical Vlasov simulations of the Hamiltonian Mean-Field model. This model is used as a representative of the class of systems under long-range interactions. We check existing results on the stability of the homogeneous situation and analyze numerical properties of the semi-Lagrangian time-split algorithm for solving the Vlasov equation. We also detail limitations due to finite resolution of the method.
 \end{abstract}
 \end{frontmatter}

\section{Introduction}

Numerical resolution of the Vlasov equation plays a crucial role in the field of plasma physics, as a theoretical investigation tool or to study realistic plasma devices. Development of Vlasov codes is still an active research topic \cite{shoucri_eulerian_codes_2008}, presenting technical aspects (parallelization, memory usage limitations, performing full 6D simulations \cite{crouseilles_patches_2007}) as well as theoretical ones (numerical dissipation, ``non-Vlasov'' effects \cite{galeotti_califano_prl_2005,califano_galeotti_pop_2006}).
Recently, the Vlasov framework has been used to study toy models of particles under long-range interactions \cite{yamaguchi_et_al_physica_a_2004,chavanis_hmf_epjb_2006,antoniazzi_califano_prl,antoniazzi_prl}. These models comport an all to all coupling and model more complex physical situations such as Coulomb or gravitational interactions. A peculiarly slow relaxation to thermodynamical equilibrium has been discovered in these systems, and understood with the help of the Vlasov equation.

A model displaying many phenomena related to long-range interactions is the Hamiltonian Mean-Field model (HMF) \cite{antoni_ruffo_1995}. It displays an equilibrium phase-transition from an homogeneous to a non-homogeneous state, characterized by a mean-field order parameter. Depending on the initial condition, the HMF experiences an out-of-equilibrium phase transition.
A theory developed by Lynden-Bell for galactic collisionless dynamics predicts the coarse-grained asymptotic evolution of the particles' distribution. It has been succesfully applied to the HMF, predicting the out-of-equilibrium phase diagram \cite{antoniazzi_prl,chavanis_hmf_epjb_2006}.

Only one study related to the HMF makes use of a Vlasov simulation~\cite{antoniazzi_califano_prl}. A good knowledge of the numerical procedure, applied to models with long-range interactions, is however necessary if one aims at widening the toolset used to study their dynamical properties, especially the concept of the so-called Quasi-Stationary States that depend on the number of particles considered. The other way round, the use of simple models in Vlasov simulation may offer a better analysis of the tool itself, serving the needs of plasma physics research.

We present in this article detailed numerical simulations of the Vlasov equation associated to the HMF model. We check existing results obtained via N-body simulations and analyze the numerical aspects of the simulations.

Section \ref{sec:system} presents the HMF, section \ref{sec:algo} presents the algorithm used. We present the results of the computations in section \ref{sec:results} and conclude in section \ref{sec:conclu}.

\section{Modeled system}
\label{sec:system}
The Hamiltonian Mean-Field (HMF), introduced in Ref. \cite{antoni_ruffo_1995}, consists of $N$ rotators interacting with a cosine attractive potential. It provides a simplified model of self-gravitating sheets, its potential corresponding to the first Fourier mode of the latter model. The Hamiltonian is~:
\begin{equation}
  \label{eq:HMF}
  H = \sum_{i=1}^N \frac{p_i^2}{2} + \frac{1}{2 N} \sum_{i,j=1}^N \left( 1 - \cos(\theta_i - \theta_j) \right) ,
\end{equation}
we define the vector~:
\begin{equation}
  \label{eq:M}
  \mathbf{M} = \unsurn \sum_i \mathbf{m}_i =  \unsurn \sum_i \left( \cos\theta_i\ ,\ \sin\theta_i \right) .
\end{equation}
$M = |\mathbf{M}| $ is called the magnetization and quantifies the degree of inhomogeneity of the particles; it also gives a simplified expression for the interaction part of the Hamiltonian, $V = \undemi (1-M^2)$. We shall make use of $M$ to follow the time evolution of the system.

In the mean-field limit $N\to\infty$, the HMF is described by a Vlasov equation~:
\begin{eqnarray}
  \label{eq:vlasov}
  \frac{\partial f}{\partial t} &+& p \frac{\partial f}{\partial \theta} - \frac{dV[f]}{d\theta} \frac{\partial f}{\partial p} = 0 \cr
    & & \cr
  V[f](\theta) &=& 1 - M_x[f] \cos\theta - M_y[f] \sin\theta \cr
  M_x[f] &=& \int d\theta dp\ f \cos\theta, \cr
  M_x[f] &=& \int d\theta dp\ f \sin\theta
\end{eqnarray}
where $f=f(\theta,p ; t)$ is the one-particle distribution function,  $V$ is the potential, depending self-consistently of $f$. The energy is conserved, and takes the form
\begin{equation}
  \label{eq:U}
  U(t)[f] = \int d\theta\ dp\ f(\theta,p ; t) \left( \frac{p^2}{2} + \undemi \left( 1 - M_x[f]\cos\theta - M_y[f]\sin\theta \right) \right) .
\end{equation}

In the following, we make use of homogeneous initial conditions, $f(\theta,p)=f(p)$ and study their consequent dynamics.
An homogeneous state is by definition stationary because $M_x$ and $M_y$ (hence the force-field) are zero, but may be stable or unstable. A stability criterion has been devised by Yamaguchi {\it et al} \cite{yamaguchi_et_al_physica_a_2004} (see also Chavanis \cite{chavanis_hmf_epjb_2006}). Waterbag and Gaussian distributions display an energy treshold below which an unstability occurs, giving rise to a magnetized state; N-body simulations were performed in \cite{yamaguchi_et_al_physica_a_2004} to check these predictions. The magnetized state displays oscillations of $M$ around a value that is found to be below thermodynamical equilibrium. Finite N effects eventually allow the system to reach equilibrium. 

Lynden-Bell has devised a statistical theory to predict the outcome of the collisionless galactic dynamics. It has been successfully applied to the HMF, in the case of waterbag initial conditions \cite{antoniazzi_prl}, and compared to Vlasov simulations \cite{antoniazzi_califano_prl}. It predicts a $f$ that is different from equilibrium; this solution is interpreted as an intermediary stage of the N-body dynamics, hence the term Quasi-Stationary State, whose lifetime diverges with $N$.

\section{Algorithm}
\label{sec:algo}

We use the algorithm described in Ref. \cite{cheng_knorr_1976}, reviewed with cubic spline interpolation \cite{sonnendrucker_et_al_semi-lag_1999}. It is based on the characteristics method for solving a partial differential equation : the value of $f(\theta,p)$ at time $t+\Delta t$ is taken from $f$ at time $t$, at the foot of the characteristic curve:
\begin{equation}
  \label{eq:CandK}
  f^{s+1}(\theta,p) = f^s( \theta- \Delta t (p + \frac{1}{2} F^\ast(\bar \theta)\Delta t) , p - \Delta t F^\ast(\bar \theta))
\end{equation}
where $\bar \theta = \theta - p\Delta t /2$ and $F^\ast$ is the force field computed at half a time-step. We use the notation $f^s(\theta,p) = f(\theta,p ; s\ \Delta{}t)$.

A practical implementation to compute Eq. (\ref{eq:CandK}) on a numerical mesh is the following :

\begin{tabular}{l l l}
  1. & Advection in the & $f^\ast(\theta_i,p_m) = f^s(\theta_i-p_m\Delta t/2, p_m)$\cr
     & $\theta$-direction, 1/2 time-step & \cr
  2. & Computation of the force field & \cr
     & for $f^\ast$ & \cr
  3. & Advection in the & $f^{\ast\ast}(\theta_i,p_m) = f^\ast(\theta_i, p_m - F^\ast(\theta_i) \Delta t)$\cr
     & $p$-direction, 1 time-step & \cr
  4. & Advection in the & $f^{s+1}(\theta_i,p_m) = f^{\ast\ast}(\theta_i-p_m\Delta t/2, p_m)$\cr
     & $\theta$-direction, 1/2 time-step & 
\end{tabular}

\begin{figure}[h]
  \centering
  \includegraphics[width=\linewidth]{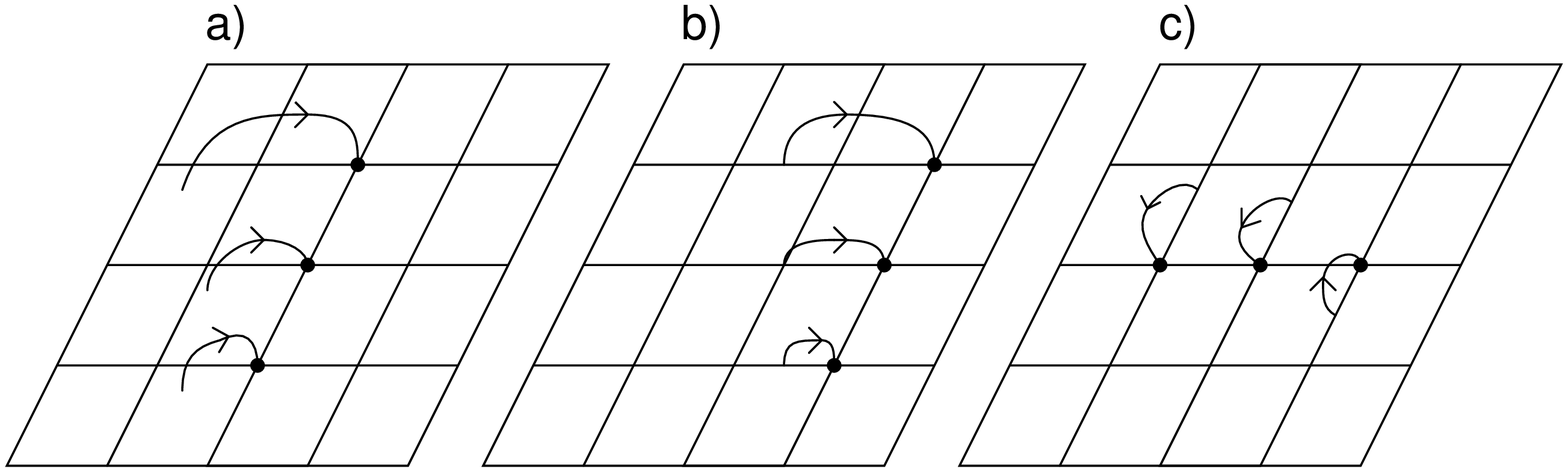}
  \caption{The semi-Lagrangian algorithm: a) illustrates the computation of the value of $f$ at the grid points indicated by the $\bullet$'s from the feet of the characteristic curves (the start of the arrow). b) and c) represent steps 1 and 3 of the time-split algorithm.}
  \label{fig:grid_all}
\end{figure}

Each of these steps is performed for the whole grid before going to the next one. A sketch of these steps is given in Fig. \ref{fig:grid_all}. This method, involving a fixed mesh and trajectories along the characteristics backwards in time is called semi-Lagrangian. It is used in fluid dynamics, for weather forecasts for instance (see Ref. \cite{staniforth_cote_1991}).

The time-split approach is valid at the second order for the Vlasov equation \cite{cheng_knorr_1976}. An interpolation scheme must be provided in order to compute $f^s(\theta,p)$ outside of the storage mesh; we will use cubic spline interpolation \cite{NR_in_f90}. The interpolation needs only to be 1-dimensional, whether in the $\theta$-space or in the $p$-space, simplifying the numerical procedure. Cubic splines are commonly used in plasma simulations, but other advection algorithms exists that can improve conservation or monotonicity properties, depending on the situation, see Refs. \cite{filbet_et_al_conserv_schemes_2001,arber_vann_critical_compar_2002}.

Altough the Vlasov equation conserves theoretically every invariant of the form $C_s[f] = \int d\theta\ dp\ s(f(\theta,p))$ (the so-called Casimirs), the numerical scheme shows deviations. We define the $L_i(t)$ norms~:
\begin{equation}
  \label{eq:lnorm}
  L_i(t) = \int d\theta\ dp \left( f(\theta,p) \right)^i ,
\end{equation}
$L_1$ is the mass of the particles (it is normalized to $1$); we will use $L_2$ to characterize numerical dissipation.

The computations were performed in Fortran~90, and the data read with the help of numpy, h5py and matplotlib~\cite{oliphant_cise_2007,mpl,h5py}.

\section{Results}
\label{sec:results}

We discuss in this section the computations related to the properties of the simulation for an initial gaussian homogeneous distribution, then we check the stability criterion of Ref. \cite{yamaguchi_et_al_physica_a_2004}. We also compare the runs in the spirit of Refs. \cite{galeotti_califano_prl_2005,califano_galeotti_pop_2006}.

\subsection{Properties of the numerical solution}
\label{sec:prop}

We run simulations starting with a gaussian profile, slighty perturbed~:
\begin{equation}
  \label{eq:IC_G}
  f(\theta, p) = \sqrt{\frac{\beta}{2\pi}} e^{-\beta p^2/2} \left( 1 + \epsilon \sin\theta \right)
\end{equation}
at an energy of $U=0.51$, $\epsilon=10^{-4}$. $\beta$ is computed from $U$, for an homogeneous state~: $\beta=\undemi\frac{1}{U-1/2}$. At $U=0.51$, the gaussian profile is unstable; $M$ grows, then oscillates around a mean value to which it eventually relaxes.

The following parameters are used : $\Delta t = 0.1$, size of the box in the $p$-direction $[-4.5:4.5]$. The resolution is varied from $N_\theta=N_p=64$ (G64) to $N_\theta=N_p=512$ (G512); we call these runs G64, G128, G256 and G512.

We detail in Table \ref{tab:cons} the conservation properties of the algorithm. The conservation of $U$ could be improved by decreasing $\Delta t$, implying more computational time. This is the sole constraint on $\Delta t$ as the semi-Lagrangian method has no Courant condition.

\begin{table}[h]
  \centering
  \begin{tabular}{l | r | r}
    run & $ \max \frac{L_1(t)-L_1(0)}{L_1(0)} $ & $ \max \frac{U(t)-U(0)}{U(0)}$ \cr
    \hline\hline
    & & \cr
    G64  & $1.6\ 10^{-4}$ & $3.2\ 10^{-4}$ \cr
    G128 & $2.1\ 10^{-5}$ & $2.0\ 10^{-4}$ \cr
    G256 & $7.2\ 10^{-6}$ & $2.0\ 10^{-4}$ \cr
    G512 & $2.8\ 10^{-6}$ & $2.0\ 10^{-4}$
  \end{tabular}
  \caption{Conservation properties at different resolutions.}
  \label{tab:cons}
\end{table}

\begin{figure}[h]
  \centering
  \begin{minipage}[t]{.49\linewidth}
  \includegraphics[width=\linewidth]{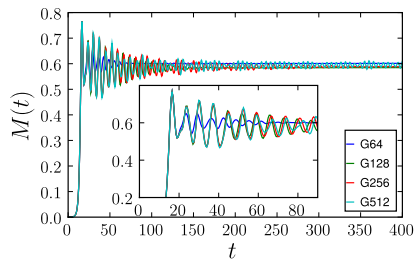}
  \caption{Evolution of $M$ for the gaussian initial condition, Eq. (\ref{eq:IC_G}) ($U=0.51$,$\epsilon=10^{-4}$). A magnification of the same curve is presented in the inset.}
  \label{fig:G_M}    
  \end{minipage}
  \begin{minipage}[t]{.49\linewidth}
  \includegraphics[width=\linewidth]{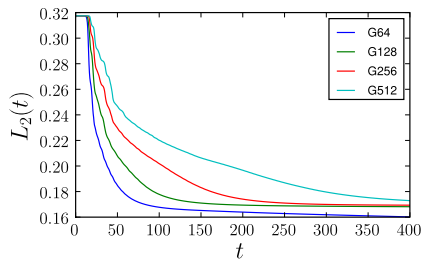}
  \caption{Evolution of $L_2$, same runs as Fig.~\ref{fig:G_M}.}
  \label{fig:G_I2}    
  \end{minipage}
\end{figure}

$M$ displays a similar behaviour for all values of the resolution until $t\approx 60$, except for the lowest resolution run G64; $M$ has been checked to correspond to N-body dynamics with $N=10^6$ for short times (data not shown).
$M$ grows up to a saturation value, identical for all runs, then oscillations take place and eventually damp to a stable value.

\begin{figure}[h]
  \centering
  \includegraphics[width=\linewidth]{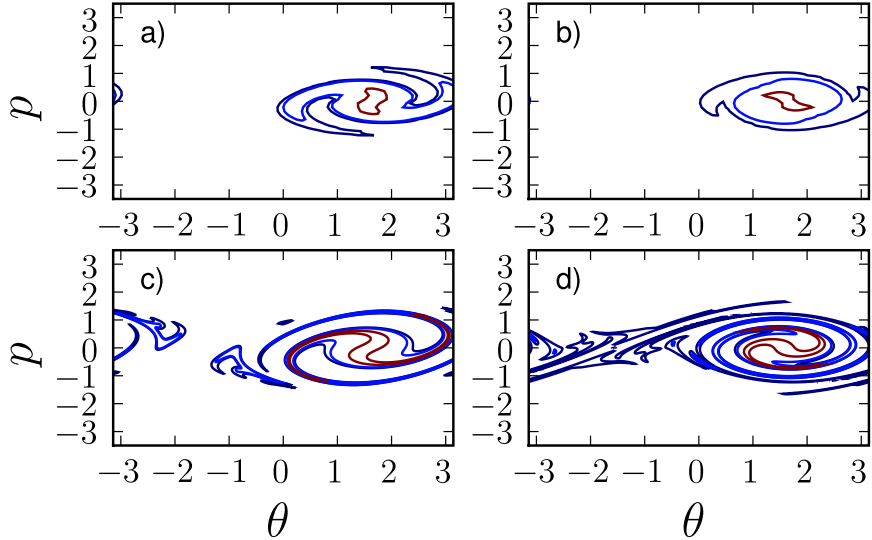}
  \caption{Phase space contour lines. a) and b) are from run G64 at times 32 and 64. c) and d) are from run G512 at the same times. While a) still presents the correct shape as compared to c), b) has lost most of the spiral shape present in d).}
  \label{fig:G_f}
\end{figure}

The damping of the oscillations is strongest in G64, they disappear at $t\approx70$. This damping is present in the other runs as well, a phenomenon that diminishes when the resolution is increased. This fact is to be put in correlation with the evolution of $L_2$ (Fig. \ref{fig:G_I2})~: The decrease of $L_2$ means that the small scales are not well described anymore by the numerical procedure. This decrease is the strongest in G64 where $L_2$ doesn't reach a plateau and diminishes continuously.
The shape of $f$, a vortex rotating in phase space, is giving the oscillations of $M$. When this structure is smoothed out by the lost of the small scales, see Fig. \ref{fig:G_f}, $M$ tends to a constant value. Run G512 displays a filamentary structure and oscillations in $M$ up to the end of the simulation.

Asymptotically, all runs tend to a similar value of $M$, which we understand with the standard deviation of $M$, $\sigma_M$, and the mean value across simulations, $\bar M$. $\sigma_M/\bar M = 1.1\%$, or $0.7\%$ if we leave G64 aside.

\subsection{Stability of the initial condition}

The stability of homogeneous initial condition has been studied in \cite{yamaguchi_et_al_physica_a_2004}. We perform a check for the waterbag and gaussian initial profiles. The waterbag is the following profile, for a given $U$:
\begin{equation}
  \label{eq:IC_WB}
  f(\theta,p) = \left\{
    \begin{array}{r l}
      \frac{1}{4 \pi \sqrt{6U}}, & p \leq \sqrt{6U}\cr
      0,                        & \mbox{else.}
    \end{array}\right.
\end{equation}

\begin{figure}[h]
  \centering
  \begin{minipage}[t]{.49\linewidth}
    \includegraphics[width=\linewidth]{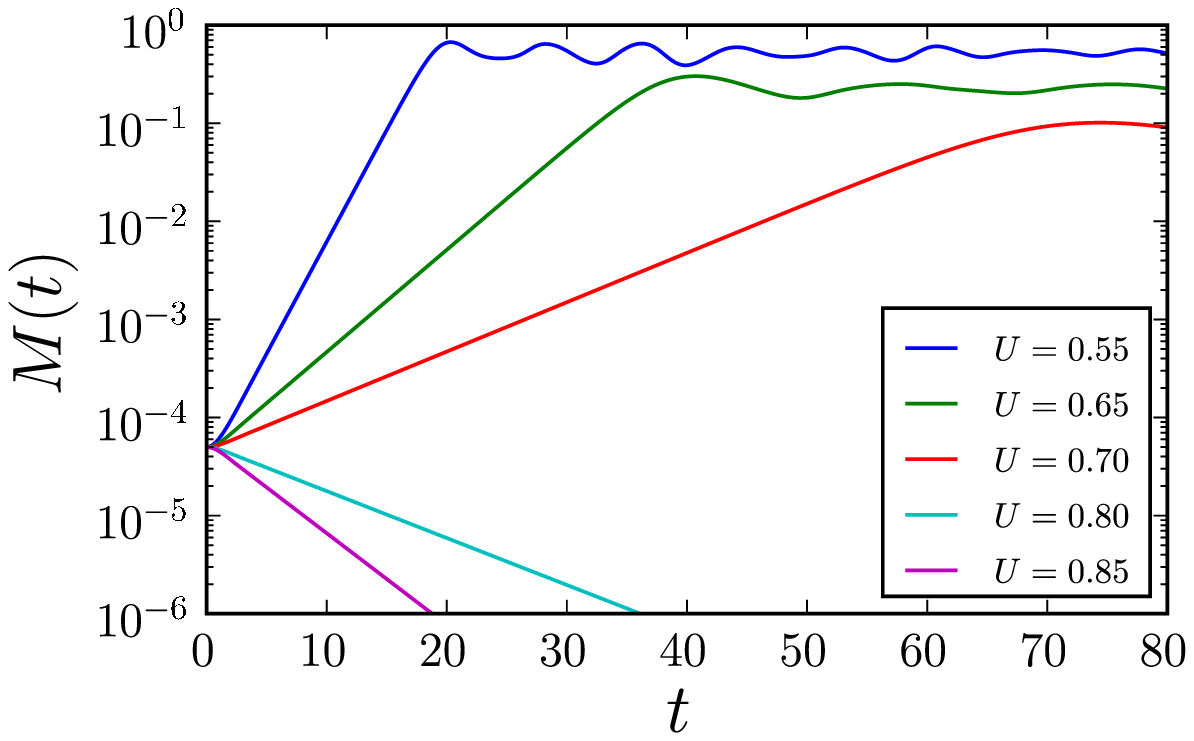}
    \caption{$M(t)$ for initial gaussian profiles. $M$ grows if $U$ is below $U_c^\ast=3/4$.}
    \label{fig:SG_M}
  \end{minipage}
  \begin{minipage}[t]{.49\linewidth}
    \includegraphics[width=\linewidth]{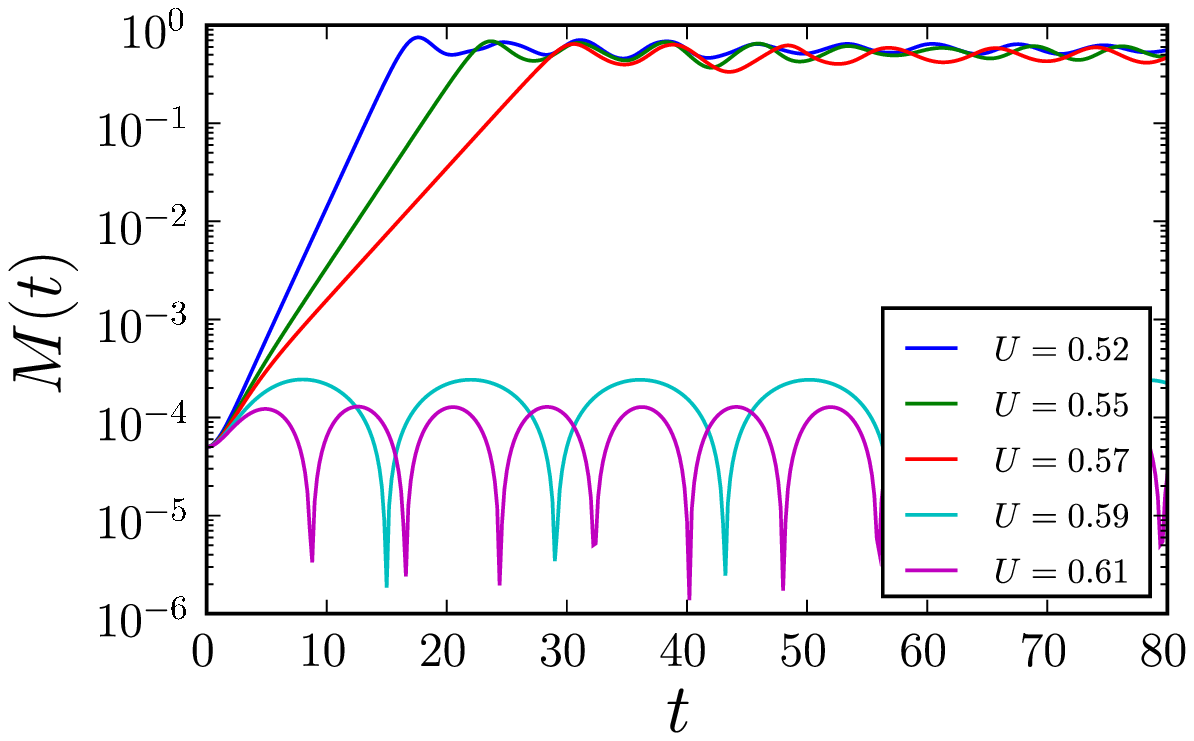}
    \caption{$M(t)$ for initial waterbag profiles. $M$ grows if $U$ is below $U_c^\ast=7/12\approx 0.58$. The slope before saturation of the three first runs is used to compute the exponential growth factor $\lambda$.}
    \label{fig:SW_M}
  \end{minipage}
\end{figure}

Starting from both side of the critical energy ($U_c^\ast=7/12$ for the waterbag and $U_c^\ast=3/4$ for the gaussian), we observe in Figs. \ref{fig:SG_M} and \ref{fig:SW_M} the change in stability. For values of $U$ above $U_c^\ast$, $M$ stays at the same order of magnitude or decreases.

In the case of the waterbag, we compared the exponential growth rate defined in Ref. \cite{yamaguchi_et_al_physica_a_2004} with the one measured from the runs of Fig. \ref{fig:SW_M} and found a very good agreement. The exponential growth rate depends on the energy $U$~: $\lambda = \sqrt{6\left(U_c^\ast -U\right)}$. The theoretical and numerically computed values are given in couples $(\lambda_\textrm{th},\lambda_\textrm{num})$ for $U=0.52,0.55 \textrm{ and } 0.57$ respectively : $(0.61,0.61)$, $(0.43,0.43)$ and $(0.32,0.31)$.

\subsection{Effect of the resolution}

\begin{figure}[h]
  \centering
  \begin{minipage}[t]{.49\linewidth}
    \includegraphics[width=\linewidth]{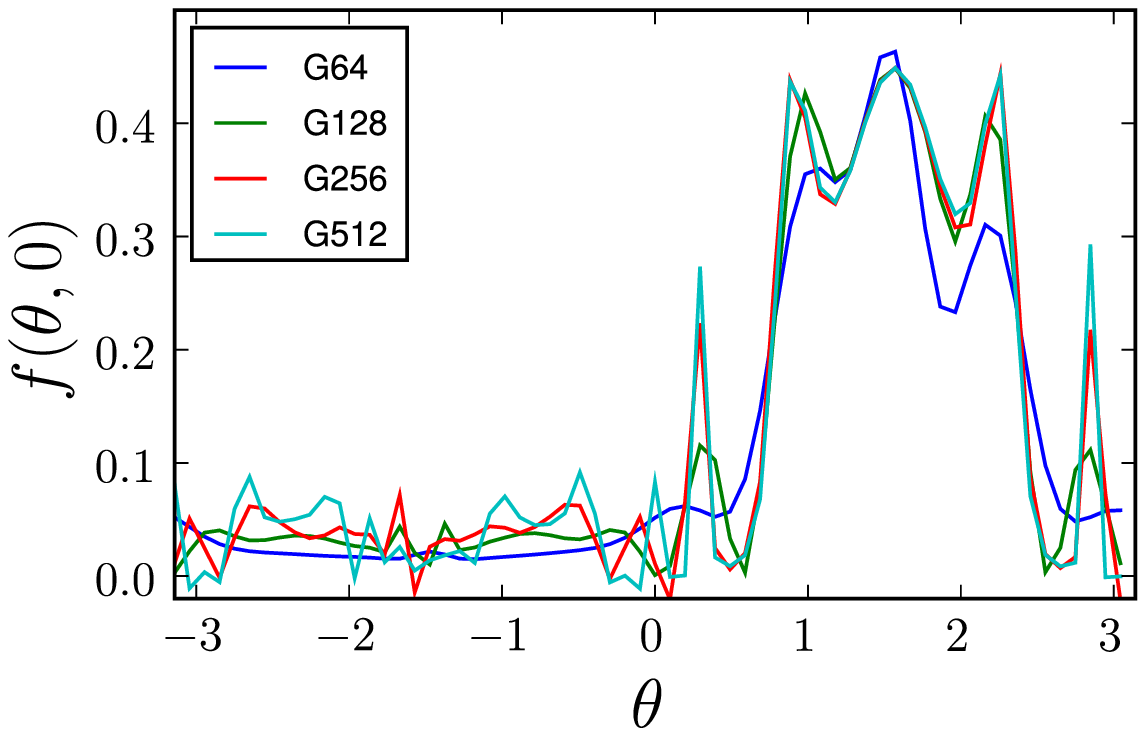}
    \caption{Cut in the function $f$ at time $t=64$. Same runs as Fig.~\ref{fig:G_M}.The locations of the peaks is similar in all runs.}
    \label{fig:G_fcut_64}
  \end{minipage}
  \begin{minipage}[t]{.49\linewidth}
    \includegraphics[width=\linewidth]{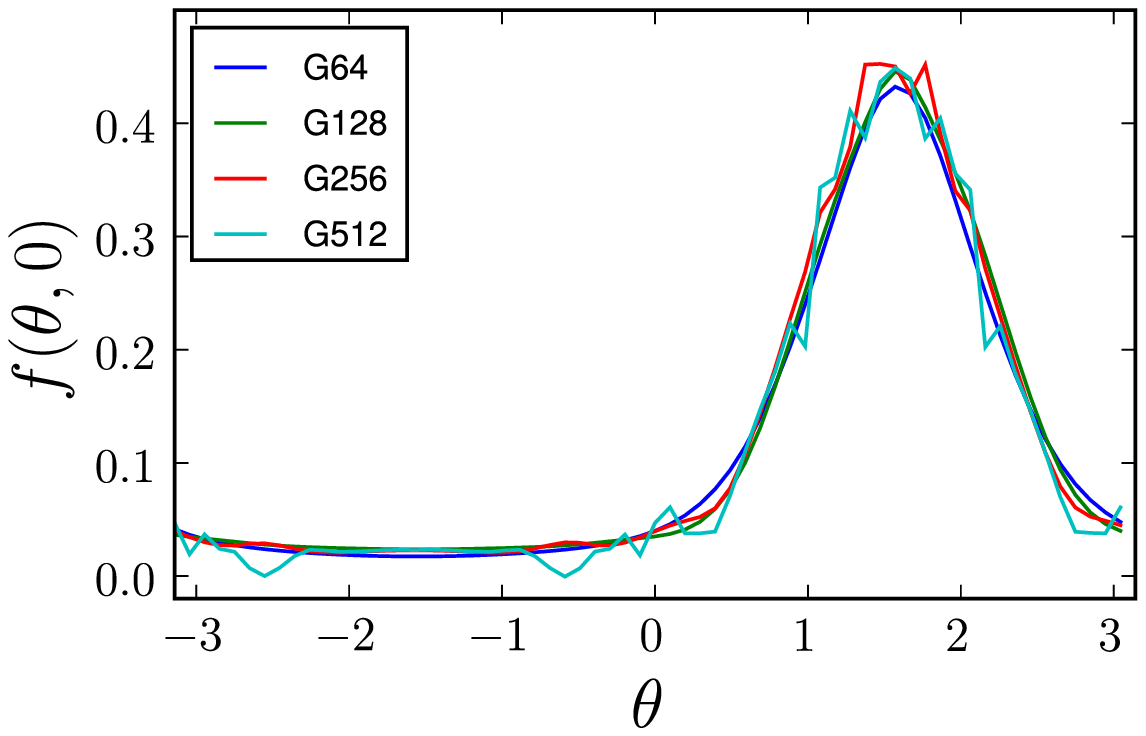}
    \caption{Cut in the function $f$ at time $t=400$. Same runs as Fig.~\ref{fig:G_M}. The location of the maximum is the same in this asymptotic state.}
    \label{fig:G_fcut_400}
  \end{minipage}
\end{figure}

The results of section \ref{sec:prop} considered different resolutions of the same problem; $f(\theta,p)$ was inspected visually, but most of the analysis focused on macroscopic quantities : $L_1$, $L_2$, $U$ and $M$. Galeotti, Califano and Mangeney \cite{galeotti_califano_prl_2005,califano_galeotti_pop_2006} compared cuts in $f$, i.e. $f(\theta,p^\ast)$ where $p^\ast$ is fixed or $f(\theta^\ast,p)$ where $\theta^\ast$ is fixed, after smoothing of all runs to the lowest resolution. This method explores with more precision the structure of $f$. For runs G64 to G512, we display $f(\theta,0)$ in Figs. \ref{fig:G_fcut_64} and \ref{fig:G_fcut_400} at two times~: $t=64$ when runs G128 to G512 still have a similar $M$, and at $t=400$ in order to discuss the asymptotic evolution.

At $t=64$, altough $L_2$ values are different, which may imply a different evolution of $f$, we find peaks in $f(\theta,0)$ at the same locations. This indicates that $f$ behaves similarly. Indeed, the energy spectrum of the HMF has only one component and the force on the particles (or phase space elements) only depends on $M$. Thus, as long as $M$ is identical for the runs, so is the force term.

At $t=400$, G64 is completely smooth, and the other runs still show a small filamentary structure. The location of the maximum is the same, a fact that contrasts with the results of Ref. \cite{califano_galeotti_pop_2006} for a Vlasov plasma, and that indicates a similar asymptotic state.

We gain by these observations, for this model, the fact that one observable, $M$, not only describes the macroscopic evolution, but by comparison between runs also monitors the structure of $f$. Development of Vlasov codes may benefit from this simplicity while comparing different algorithm, a process often used in numerical analysis (for Vlasov plasmas, see for instance Ref. \cite{arber_vann_critical_compar_2002}).

\section{Conclusions}
\label{sec:conclu}

We presented in this paper a detailed study of Vlasov numerical simulations for the Hamiltonian Mean-Field model (HMF), that have been compared to previous work on the HMF {and} to Vlasov results in plasma physics. We gave evidence that the mean-field order parameter $M$ of the HMF is sufficient to compare different runs in terms of phase space evolution as well as dissipative (i.e. damping) properties.

Illustration was given of the ``non-Vlasov'' limit of the numerical simulation which we need to keep in mind to analyze simulations results carefully, as stated recently by other authors \cite{galeotti_califano_prl_2005,califano_galeotti_pop_2006,antoniazzi_califano_prl}.

Finally, we hope that cross research between long-range systems obeying a Vlasov equation and Vlasov plasmas may benefit to both fields by providing the first with a new tool they can apprehend and the latter with new insight on the numerical method.

\section*{Acknowledgments}

The author would like to thank P. Gaspard and M. Malek-Mansour for their support. The author acknowledges useful discussions with D. Fanelli and comments from J.-S. McEwen. This research is financially supported by the Belgian Federal Government (Interuniversity Attraction Pole ``Nonlinear systems, stochastic processes, and statistical mechanics'', 2007-2011).

\end{document}